# HEBS: Histogram Equalization for Backlight Scaling

**Ali Iranli, Hanif Fatemi, Massoud Pedram**
Department of Electrical Engineering
University of Southern California
*{iranli,fatemi,pedram}@usc.edu*

***ABSTRACT*** - *In this paper, a method is proposed for finding a pixel transformation function that maximizes backlight dimming while maintaining a pre-specified image distortion level for a liquid crystal display. This is achieved by finding a pixel transformation function, which maps the original image histogram to a new histogram with lower dynamic range. Next the contrast of the transformed image is enhanced so as to compensate for brightness loss that would arise from backlight dimming. The proposed approach relies on an accurate definition of the image distortion which takes into account both the pixel value differences and a model of the human visual system and is amenable to highly efficient hardware realization. Experimental results show that the histogram equalization for backlight scaling method results in about 45% power saving with an effective distortion rate of 5% and 65% power saving for a 20% distortion rate. This is significantly higher power savings compared to previously reported backlight dimming approaches.*

## 1 Introduction

As the portable electronic devices become more intertwined with everyday life of people, it becomes necessary to put more functionality into these devices, run them at higher circuit speeds, and have them consume a small amount of energy. These electronic devices are becoming smaller and lighter and often required to operate with fancy Liquid Crystal Displays (LCD's) for increasing periods of time. Unfortunately, the battery capacities are increasing in much slower pace than the overall power dissipation of these kinds of devices. Therefore, it is essential to develop low power design techniques to reduce the overall power dissipation of these devices.

Previous studies on battery powered electronic devices have pointed out that the energy consumption in the Cold Cathode Fluorescent Lamp (CCFL), which is the backlight of an LCD, dominates the overall energy consumption of the device [1]. In the SmartBadge system for instance, the display consumes 28.6%, 28.6%, and 50% of the total power in the active, idle, and standby modes, respectively [1]. Unfortunately, as for other resources, one cannot tackle this energy hungry component with some form of power shutdown. This is due to the fact that LCD subsystem must be continuously refreshed and cannot be turned off or put to sleep without a significant penalty in performance and Quality of Service (QoS).

There are two main classes of techniques for lowering the power consumption of LCD subsystem. The first class of techniques is focused on the digital/analog interface between the graphics controller and the LCD controller [2][3]. These techniques try to minimize the energy consumption by taking advantage of different encoding schemes to minimize the switching activity of the electrical bus. For instance, reference [2] uses the spatial locality of the video data to reduce the number of transition on the DVI bus reducing its energy consumption by 75%. More recently, reference [3] has extended the previous work by using the limited intra-word transition codes instead of single intra-word transition codes used in [2] and including the DC balancing conditions in the coding scheme to achieve more than 60% energy saving, on average compared to the basic transmission protocol. The second class of techniques is focused on the video controller and the backlight to lower the energy consumption of the display system. The key idea of these techniques follows from the observation that eye's perception of the light, which is emitted from the LCD panel, is a function of two parameters, 1) the light intensity of the backlight and 2) the transmittance of the LCD panel. Therefore, by carefully adjusting these two parameters one can achieve the same perception in human eyes at different values of the backlight intensity and the LCD transmittance. However, since the changes in energy consumption of the backlight lamp is higher than that of the LCD panel by orders of magnitude, one can save energy by simply dimming the backlight and increasing the LCD transmittance to compensate for the loss of backlight.

Reference [4] was the first to propose a simple approach to the backlight scaling problem, where each pixel value is changed to achieve a higher transmittance through the LCD panel, thereby creating the possibility to reduce the backlight intensity and hence lowering the energy consumption. The main drawbacks of this approach are twofold; a) there is no measure of how distorted the new image would become, and b) manipulating pixels one by one is a time consuming process that limits the applicability of this approach. *Image distortion* after backlight luminance dimming is evaluated by the percentage of saturated pixels that exceed the range of pixel values, e.g., [0..255]. In contrast, reference [5] improved this basic approach by proposing the *contrast fidelity* as a measure of image distortion and solving the optimal transmittance scaling policy problem. Moreover, an elegant approach for hardware implementation of the transmissivity scaling function by using the built-in reference voltages is proposed which eliminates the pixel-by-pixel manipulation of the image. The authors report a 3.7X power saving with less than 10% contrast distortion in the image.

In this paper, a method is proposed for finding a pixel transformation function that maximizes backlight dimming while maintaining a pre-specified distortion level. This is achieved with the following algorithm, called HEBS (for *Histogram Equalization for Backlight Scaling*):

1) Given an original image, $\chi$, and an upper bound on the tolerable image distortion, we determine the minimum dynamic range, $R$, of pixel values in a transformed image. The goal is to achieve the maximum power savings as a result of a follow-on backlight dimming. Therefore, this step also produces the optimum backlight scaling factor, $\beta$.
2) We determine a transformation function, $\Phi$, which takes the original image histogram (with dynamic range $N$) to a uniform distribution histogram with a range of $R$. This is in turn accomplished by mapping the $N$ pixel values in the original image to the $R$ target values of the transformed image so that the difference between the resulting histogram and the uniform distribution histogram is minimized.
3) For reasons having to do with the existing infrastructure for grayscale level assignment and the efficiency of a hardware-based implementation of the proposed transformation, we approximate $\Phi$ with a piecewise linear function, $\Lambda$, such that the squared error difference between $\Phi$ and $\Lambda$ is minimized.
4) We construct the transformed image by applying $\Lambda$ to the original image, $\chi$. At the same time, we dim the backlight by factor $\beta$.

The advantages of HEBS compared to previous backlight dimming techniques are:

1) A more accurate definition of the image distortion which takes into account both the pixel value differences and a model of the human visual system is used.
2) A powerful image histogram "compression" technique which allows us to reduce the dynamic range of an image subject to a given image distortion level. This should be contrasted to prior works which can only reduce the image dynamic range by saturating the pixel values either at one end [4] or both ends [5] of the image histogram.
3) Amenable to highly efficient hardware realization by making minimal change to the existing Programmable LCD Reference Driver (PLRD).
4) An additional power saving of 15% compared to the best of the existing strategies. In a typical battery-powered mobile computer system, this constitutes a total additional system power saving of 3% in active mode.

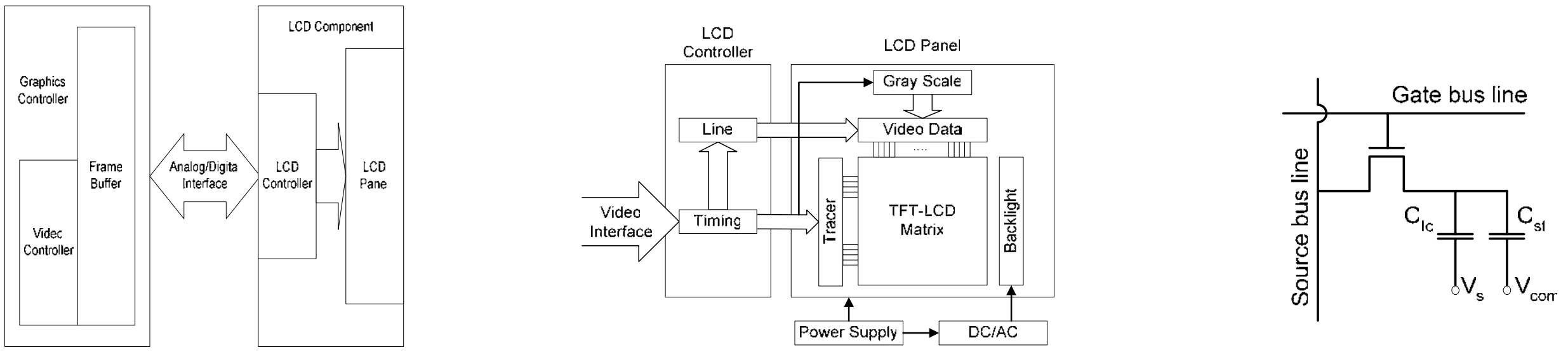

a. LCD subsystem architecture  b. LCD component architecture  c. TFT cell schematic

Figure 1. TFT-LCD screen

## 2 Background

Figure 1a shows the typical architecture of the digital LCD subsystem in a microelectronic device. There are two main components in this subsystem: a) video controller and frame buffer memory and b) LCD controller and panel. The image data, which is received from the processing unit, is first saved into the frame buffer memory by the video controller and is subsequently transmitted to the LCD controller through an appropriate analog (e.g., VGA) or digital (e.g., DVI) interface. The LCD controller receives the video data and generates a proper grayscale – i.e., transmissivity of the panel – for each pixel based on its pixel value. All of the pixels on a transmissive LCD panel are illuminated from behind by the backlight. To the observer, a displayed pixel looks bright if its transmittance is high (i.e., it is in the 'on' state), meaning it passes the backlight. On the other hand, a displayed pixel looks dark if its transmittance is low (i.e., it is in the 'off' state), meaning that it blocks the backlight. For color LCD's, different filters are used to generate shades of three main colors (i.e. red, blue, and green), and then color pixels are generated by mixing three sub-pixels together to produce different colors.

Figure 1b depicts the LCD controller in more detail. The data received from the video bus is used to infer a row of pixels timing information and respective grayscale levels. Then, this timing information is used to select the appropriate row in the LCD matrix. Next, the pixel values are converted to the corresponding voltage levels to drive the thin-film-transistors (TFT's) on different columns of the selected row. The backlight bulb is powered with the aid of a DC-AC converter, to provide the required illumination of the LCD matrix.

Each pixel has an individual liquid crystal cell, a TFT, and a storage capacitor (cf. Figure 1c). The electrical field of the capacitor controls the transmittance of the liquid crystal cell. The capacitor is charged and discharged by the TFT. The gate electrode of the TFT controls the timing for charging/discharging of the capacitor when the pixel is scanned (or addressed) by the tracer for refreshing its content. The (drain-) source electrode of the TFT controls the amount of charge. The gate electrodes and source electrodes of all TFT's are driven by a set of gate drivers and source drivers, respectively. A single gate driver (called a *gate bus line*) drives all gate electrodes of the pixels on the same row. The gate electrodes are enabled at the same time the row is traced. A single source driver (called a *source bus line*) drives all source electrodes of the pixels on the same column. The source driver supplies the desired voltage level (called *grayscale voltage*) according to the pixel value. In other words, ideally, the pixel value transmittance, $t(X)$, is a linear function of the grayscale voltage $v(X)$, which is in turn a linear function of the pixel value $X$. The transfer function of source driver which maps different pixel values, $X$, into different voltage levels, $v(X)$ is called the *grayscale-voltage function*. If there are 256 grayscales, then the source driver must be able to supply 256 different grayscale voltage levels. For the source driver to provide a wide range of grayscales, a number of *reference voltages* are required. The source driver mixes different reference voltages to obtain the desired grayscale voltages. Typically, these different reference voltages are fixed and designed as a voltage divider.

Mathematically speaking, in a transmissive TFT-LCD monitor, for a pixel with value $X$, the luminance $I(X)$ of the pixel[1] is:

$$I(X) = b.t(X) \tag{1a}$$

where $t(X)$ is the transmissivity of the TFT-LCD cell for pixel value $X$, and $b\in[0,1]$ is the (normalized) *backlight illumination factor* with $b=1$ representing the maximum backlight illumination and $b=0$ representing no backlight. Note that $t(X)$ is a linear mapping from [0,255] domain to [0,1] range. In backlight scaled TFT-LCD, $b$ is scaled down and accordingly $t(X)$ is increased to achieve the same image luminance.

Reference [4] describes two backlight luminance dimming techniques. This technique dims the backlight and compensates for the luminance loss by adjusting the grayscale of the image to increase its brightness or contrast. More precisely,

$$I(X) = \beta.t(\Phi(X,\beta)) \tag{1b}$$

where $0<\beta\leq 1$ is the *backlight scaling factor* and $\Phi(X,\beta)$ is the pixel transformation function.

Let $x$ denote the *normalized pixel value*, i.e., assuming an 8-bit color depth, $x=X/255$. The authors of [4] scale the backlight luminance by a factor of $\beta$ while increasing the pixel values from $x$ to $\Phi(x,\beta)$ by two mechanisms. Clearly, $\Phi(x,\beta)=x$ denotes the identity pixel transformation function (cf. Figure 2a.) The "backlight luminance dimming with brightness compensation" technique uses the following pixel transformation function (cf. Figure 2b):

$$\Phi(x,\beta) = \min(1, x+1-\beta) \tag{2a}$$

whereas the "backlight luminance dimming with contrast enhancement" technique uses this transformation function (cf. Figure 2c):

$$\Phi(x,\beta) = \min(1, \frac{x}{\beta}) \tag{2b}$$

In these schemes, the optimal backlight factor is determined by the backlight luminance dimming policy subject to the given distortion rate. To calculate the distortion rate, an image *histogram estimator* is required for calculating the statistics of the input image. Note that the image histogram simply denotes the marginal distribution function of the image pixel values.

Reference [5] proposes a different approach in which the pixel values in both dark and bright regions of the image are used to enable a further dimming of the backlight. The key idea is to first truncate the image histogram on both ends to obtain a smaller dynamic range for the image pixel values and then to spread out the pixel values in this range (by applying an *affine transformation*) so as to enable a more aggressive backlight dimming while maintaining the contrast fidelity of the image. The pixel transformation function is given as (cf. Figure 2d):

[1] Illuminance can be used to characterize the luminous flux emitted from a surface. Most photographic light meters measure the illuminance. In terms of visual perception, we perceive luminance. It is an approximate measure of how "bright" a surface appears when we view it from a given direction. Luminance is measured in lumens per square meter per steradian. The maximum brightness of a CRT or LCD monitor is described by luminance in its specification.

$$\Phi(x,\beta)=\begin{cases}0, & 0\le x\le g_l \\ cx+d, & g_l\le x\le g_u \\ 1, & g_u\le x\le 1\end{cases}\qquad c=\frac{1}{g_u-g_l}=\frac{1}{\beta};\; d=\frac{-g_l}{g_u-g_l} \tag{3}$$

where $(g_l,0)$ and $(g_u,1)$ are the points where $\Phi(x,\beta)=cx+d$ intersects $\Phi(x,\beta)=0$ and $\Phi(x,\beta)=1$, respectively.

The implementation is quite simple requiring minimal change to the built-in Programmable LCD Reference Driver (PLRD). Notice that PLRD allows a class of linear transformations on the backlight-scaled image, resulting in both brightness scaling and contrast scaling. Clearly, brightness and contrast are the two most important properties of any image. Recall that for transmissive LCD monitors, the brightness control changes the backlight illumination and the contrast control changes the LCD transmittance function.

The previous approaches cannot fully utilize the power saving potential of the dynamic backlight scaling scheme, due to the fact that their measure of distortion between the original and the backlight-scaled image is an overestimation. This is because these approaches simply either minimize the number of saturated pixel values [4] or maximize the number of pixel values that are preserved [5].

It is well known that image distortion (more precisely, the difference between a pair of similar images) is a complex function of the visual perception, and hence, it cannot be correctly evaluated by comparing the images pixel by pixel (i.e., calculating the root mean squared error of the corresponding pixel values) or as a whole (i.e., using the integral of the absolute value of the histogram differences) [7][8]. A correct measure of distortion should appropriately combine the mathematical difference between pixel values (or histograms) and the characteristics of the human visual system. One approach presented in [6] is to perform some transformation on both the original and new (i.e., backlight-scaled) images according to a human visual system model [9], and then, compare the transformed results using quantitative measures.

## 3 Histogram Equalization for Backlight Scaling (HEBS)

Due to the coherence of the objects within an image, pixels describing a single object generally have similar luminous intensity values [9].[2] The single-band grayscale spreading (cf. Figure2d) attempts to exploit this object coherence by enhancing the contrast within a band (sub-range) of the pixel values at the expense of the remaining ones, thereby, creating the opportunity for a more aggressive backlight dimming policy. However, a single band is used for the entire image. Intensity values lying above or below the band are mapped to the maximum and minimum possible intensity values, respectively. The result is that the

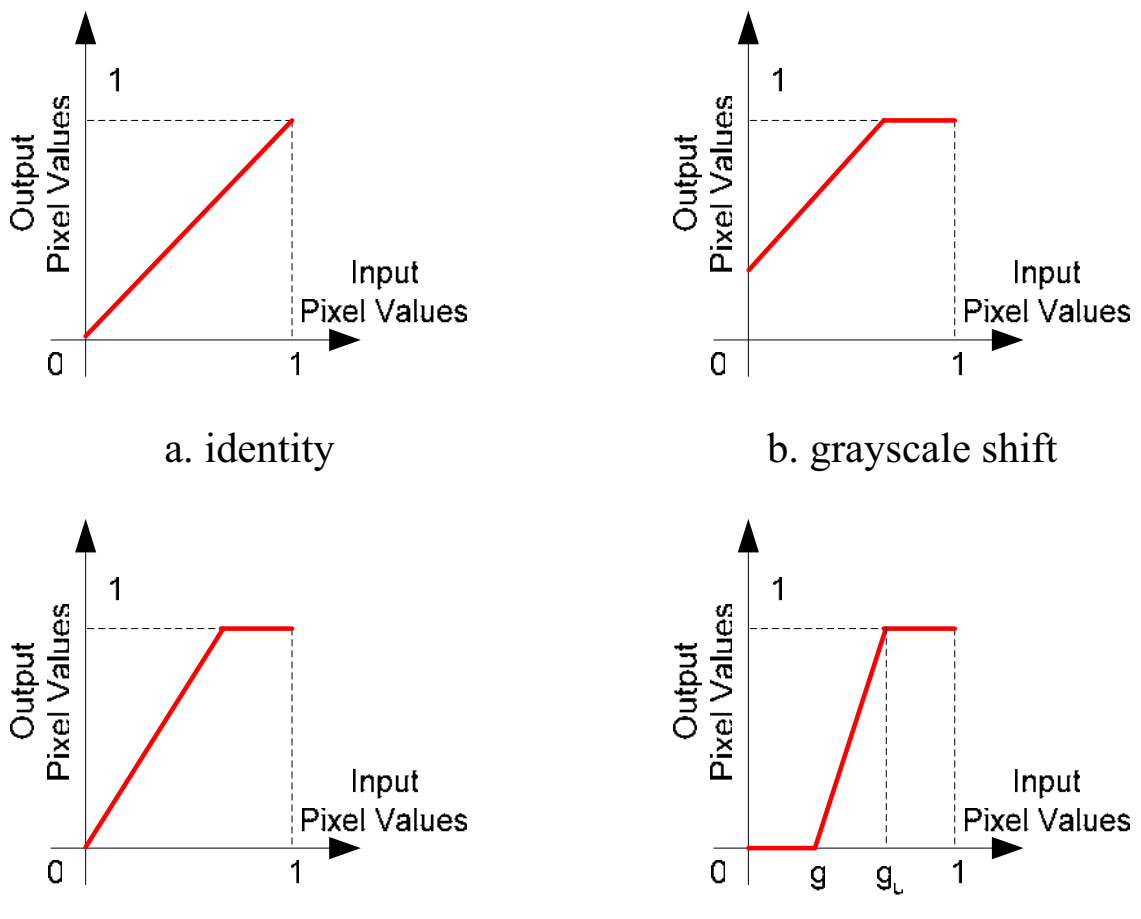

a. identity

b. grayscale shift

c. grayscale spreading

d. single-band grayscale spreading

Figure 2. Pixel transformation functions

pixel values within the window are displayed with marginal loss of quality, while information in the other parts of the image is discarded.

Let $\chi$ and $\chi'=\Phi(\chi,\beta)$ denote the original and the transformed image data, respectively. Moreover, let $D(\chi,\chi')$ and $P(\chi',\beta)$ denote the distortion of the images $\chi$ and $\chi'$ and the power consumption of the LCD-subsystem while displaying image $\chi'$ with backlight scaling factor, $\beta$.

***Dynamic Backlight Scaling (DBS) Problem:*** *Given the original image $\chi$ and the maximum tolerable image distortion $D_{max}$, find the backlight scaling factor $\beta$ and the corresponding pixel transformation function $\chi'=\Phi(\chi,\beta)$ such that $P(\chi',\beta)$ is minimized and $D(\chi,\chi')\le D_{max}$.*

The general form of DBS problem as stated above is difficult to solve due to the complexity of the distortion function, $D$, and also the non-linear function minimization step that is required to determine $\Phi(\chi,\beta)$. In the following, we will try to simplify this problem by 1) fully utilizing the dynamic range of the transformed image $\chi'$ in order to achieve the minimum TFT-LCD power consumption $P(\chi',\beta)$ and 2) by constraining the pixel transformation function to the family of piecewise linear functions (because these piecewise linear functions are desirable from implementation point of view; cf. section 4.2.)

Intuitively, to reduce the dynamic range of a given image one can discard the pixels corresponding to the grayscale levels with low population. This in turn minimizes the number of discarded pixels and hence minimizes the image distortion. On the other hand, for an image with a histogram which is uniformly populated with pixels in different grayscale levels, every level is as important as the other and discarding any grayscale level can cause a significant image distortion. Therefore, a "good" transformation which solves the DBS problem i.e., minimizes $P(\chi',\beta)$, is the one which transforms the original image histogram into a uniform intensity histogram with a minimal dynamic range. Having made this selection, we can now deal with the complexity of the distortion function as follows. We set the dynamic range of a benchmark image to some target value and plot the distortion value of the transformed image as a function of this target range We repeat this process for a number of different target ranges per image and for a large number of images in the database. Next, resorting to standard regression analysis techniques, we will calculate the "best" global fit to these distortion values. The result will be an empirical curve which maps the observed distortion function values to target dynamic range of transformed images. We call this curve the *distortion characteristic curve* (cf. section 5.1c)

## 4 Global Histogram Equalization

We propose a global histogram equalization scheme in which the intensity values in the image are altered such that the resulting image has the uniform intensity histogram, with the desired minimum ($g_{min}$) and maximum ($g_{max}$) grayscale limits. This transformation may be accomplished by the use of the cumulative distribution function of the pixel intensities to generate the intensity remapping function. In this approach the resulting image will utilize the available display levels very well, because the transformation function is based on the statistics of the entire image.

More precisely, let us denote the *cumulative* distribution histogram of the original image by $H$, and the different grayscale values of the image pixels by $x$ which are selected from a finite set of values $G$, (e.g. $G$=[0..1]). Transformation function $\Phi: G\rightarrow G$ is a monotonic function, which maps the original pixel values $x$ into a new pixel values $x'$ and as a result equalizes cumulative histogram $H$ to become the cumulative uniform histogram, $U$. (that is a sloped line going from 0 to $N$, where $N$ represents the number of pixels over which the histogram has been calculated, i.e. number of pixels in the image.)

***Global Histogram Equalization (GHE) Problem:*** *Given the original image cumulative histogram $H$, find a monotonic transformation $\Phi: G\rightarrow G$ where G=[0..1] such that*

[2] Luminous intensity is the photometric equivalent of radiance intensity.

$$\int_G \left| U(\Phi(x)) - H(x) \right| \cdot dx \tag{4}$$

*is minimized.*

If the targeted histogram is a uniform distribution between upper and lower limits, $g_{min}$ and $g_{max}$,[3] then to minimize (4) the transformation function $\Phi$ can be calculated as:

$$\Phi(x) = U^{-1}(H(x)) = g_{min} + (g_{max} - g_{min}) \cdot \frac{H(x)}{N} \tag{5}$$

In actual implementation, it is common to have a discrete version of histogram instead of the cumulative histogram. To convert (5) into a *histogram based formulation* one can differentiate both sides of (5) to get,

$$\frac{d\Phi(x)}{dx} = (g_{max} - g_{min}) \cdot \frac{h(x)}{N} \tag{6}$$

where *h(x)* denotes the *marginal* distribution histogram. We can then use the first order difference approximation for the differentiation operator, to calculate the discrete transfer function as

$$\Phi(x_i) = g_{min} + (g_{max} - g_{min}) \cdot \sum_{k=0}^{i-1} \Delta x_k \cdot \frac{h(x_k)}{N} \tag{7}$$

$$\Delta x_k \triangleq x_{k+1} - x_k$$

where $x_i \in G$ are the center points for the histogram buckets and $h(x_k)$ are the histogram value.

### 4.1 HEBS Implementation

Figure 4 depicts the flow of the HEBS algorithm. A user-specified maximum tolerable image distortion is first read as input and is subsequently used to look up the minimum admissible dynamic range for the image from the distortion characteristic curve (cf. Section 3.) Using this minimum admissible dynamic range and the transmissivity characteristics of the TFT display, a maximum backlight scaling factor, $\beta$, is calculated and used to scale down the CCFL backlight. Moreover, this minimum dynamic range along with the original image histogram will be used by the GHE problem solver to calculate the pixel transformation function $\Phi(\chi,\beta)$. Next, the transformation function is approximated by a piecewise linear function, $\Lambda(\chi,\beta)$, which is in turn used to determine the reference grayscale voltages, and to transform the original pixel values to new ones for the displayed image. In [5], Cheng et al proposed an elegant and highly efficient approach for backlight scaling implementation based on the reference voltages of the source drivers. Figure 5a shows the schematic of the built-in reference voltage driver of a typical LCD panel (recall that the voltage divider provides the required reference voltages for the source driver buffers.) For example in [11], an Analog Devices input LCD reference driver [12] is used with a 10-way voltage divider. The proposed approach in [5] adds controllable switches to both ends of the voltage driver to be able to clamp the low and high gray scale levels to the minimum and maximum gray scale values, respectively, thereby reducing the dynamic range of the image and increasing the slope of the grayscale-voltage function (cf. Sec. 2) to realize contrast compensation. However, this solution is limited in two ways; 1) It can only generate a transfer function of single-band grayscale spreading form (cf. Figure 2d), and 2) the linear region of the transfer function can only have a single slope.

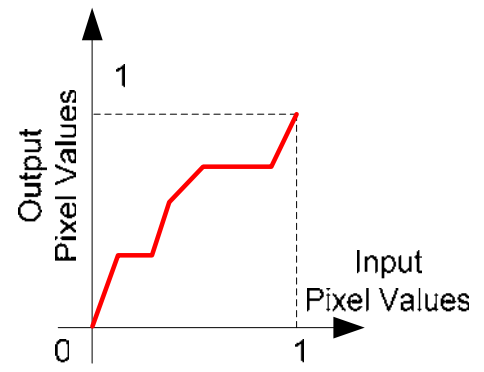

Figure 3. *k*-window grayscale spreading function

[3] $U(x)=0$ for $x < g_{min}$; $U(x)=N.(x-g_{min}) / (g_{max}-g_{min})$ for $g_{min} \le x \le g_{max}$; and $U(x)=N$ for $x>g_{max}$.

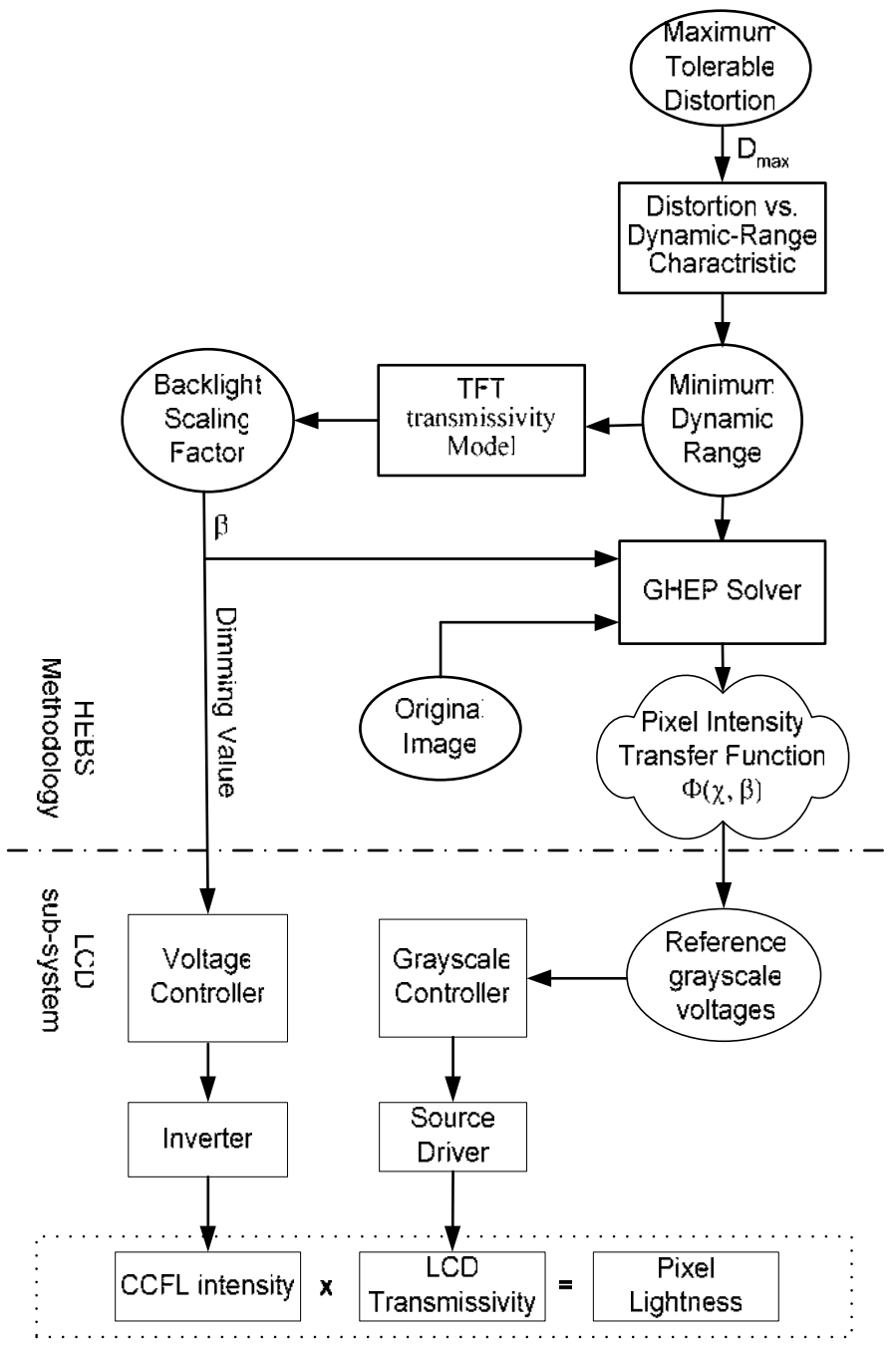

Figure 4. Histogram Equalization for Backlight Scaling flow

To implement HEBS, we use a hierarchical structure for the reference voltage dividers as shown in figure 5b. This structure provides more flexibility in creating different slopes for multiple linear regions of the grayscale-voltage transfer function. Moreover, adding switches between different grayscale levels enables one to provide flat-bands not only at the two ends of the image histogram, but also in the middle range of the gray scale levels.

To achieve multiple output slopes for the grayscale-voltage transfer function, *k* different controllable voltage sources, $V_i$, are needed (cf. Figure 5b.) These voltage sources are normally set to voltage levels $V_i = \frac{i}{k} V_{dd}$ with $i=1 \ldots k$, creating a transfer function with slope of one. Here $V_{dd}$ denotes the supply voltage, and *i* and *k* denote the voltage source number and total number of available voltage sources. To create different slopes for different regions of the grayscale values, one can change the voltage levels of controllable sources to create a *k*-band grayscale spreading function as described below. In particular we describe a procedure for approximating the pixel transformation function $\Phi(\chi,\beta)$ with a piecewise linear function $\Lambda(\chi,\beta)$, and then explain how to determine the voltage levels $V_i$, to implement this approximated function.

TFT-LCD displays are only capable of displaying a finite number of different grayscale levels, therefore, the input and output values of the transformation function $\Phi(\chi,\beta)$ are discrete. This observation implies that even the exact form of the transformation function $\Phi(\chi,\beta)$ as given by (7) is a piecewise linear function. However, the number of linear segments of $\Phi(\chi,\beta)$ is $O(|G|)$, which is too large for efficient hardware implementation. We therefore approximate $\Phi(\chi,\beta)$ with another piecewise linear function that has a small number of linear segments.

Let $P=\{p_1, \ldots, p_n\}=\{(x_1, y_1), \ldots, (x_n, y_n)\}$ denote the ordered set of endpoints of each linear segment in exact form of $\Phi(\chi,\beta)$ starting from $x_1=0$ for the darkest to $x_n=255$ for the brightest grayscale level. Moreover, let $Q=\{q_1, \ldots, q_m\}$, denote the ordered set of the endpoints of linear segments in $\Lambda(\chi,\beta)$, which is the approximation of $\Phi(\chi,\beta)$. Clearly, we have the following:

$$Q \subset P$$
$$q_1 = p_1 \text{ and } q_m = p_n \;;\; q_i = p_j \text{ and } q_{i+1} = p_k \text{ where } k > j \tag{8}$$

***Piecewise Linear Coarsening (PLC) Problem:*** *Given a piecewise linear curve P, approximate it by another piecewise linear curve Q*

*with a given number of line segments m so that the mean squared error between Φ(χ,β) and Λ(χ,β) is minimized.*

The PLC problem can be solved using a dynamic programming technique. Let *E(n,m)* denote the mean squared error between the original curve with n points and its best approximation with $m \leq n$ points. Then,

$$E(n,m) = \min_{j=m-1...n-1} (E(j,m-1) + e(j))$$
$$E(1,0) = 0,\ E(n,0) = \infty,\ and\ E(1,m) = 0 \quad \forall m,n \tag{9}$$

where *e(j)* denotes the mean squared error incurred by approximating all segments between $p_j$ and $p_n$ by a single line connecting $p_j$ to $p_n$. The time complexity of this algorithm is $O(mn^2)$.

Using the solution for the PLC problem, the voltage levels $V_i$, are calculated as

$$V_i = \frac{Y_{q_i}}{\beta} . V_{dd} \tag{10}$$

where $Y_{q_i}$ denotes the y-component of point $q_i$. Note that the backlight dimming factor β is present in denominator to spread the grayscale level of the resulting image, and hence, compensating the loss of brightness due to backlight dimming.

## 5 Experimental results

### 5.1 Characterization

**a) Cold Cathode Fluorescent Lamp (CCFL)**

The CCFL illumination is a complex function of the driving current, ambient temperature, warm-up time, lamp age, driving waveform, lamp dimensions, and reflector design [13]. In the transmissive TFT-LCD application, only the driving current is controllable. Therefore, we model the CCFL illumination as a function of the driving current only and ignore the other parameters. Accounting for the saturation phenomenon in the CCFL light source [10], we use a two-piece linear function to characterize the power consumption of CCFL as a function of the backlight factor:

$$P_{backlight}(\beta) = \begin{cases} A_{lin}.\beta + C_{lin} & 0 \leq \beta \leq C_s \\ A_{sat}.\beta + C_{sat} & C_s < \beta \leq 1 \end{cases} \tag{11}$$

Relationship between the CCFL illumination (i.e., luminous flux incident on a surface per unit area) and the driver's power dissipation for the CCFL in LG Philips transmissive TFT-LCD LP064V1 [11] is shown in Figure 6a. The CCFL illumination increases monotonically as the driving power increases from 0 to 80% of the full driving power. For values of driving power higher than this threshold, the CCFL illumination starts to saturate. The saturation phenomenon is due to the fact that the increased temperature and pressure inside the tube adversely impact the efficiency of emitting visible light [14].

After interpolation, we obtain the following coefficient values for the CCFL in LG Philips transmissive TFT-LCD LP064V1:

$C_s$=0.8234, $A_{lin}$=1.9600, $C_{lin}$=-0.2372, $A_{sat}$=6.9440, and $C_{sat}$= –4.3240.

**b) TFT-LCD display matrix**

The hydrogenated amorphous silicon (a-Si:H) is commonly used to fabricate the TFT in display applications. For a TFT-LCD panel, the a-Si:H TFT power consumption can be modeled by a quadratic function of pixel value $x \in [0,1]$ [15]

$$P_{TFT\ Panel}(x) = a.x^2 + b.x + c \tag{12}$$

We performed the current and power measurements on the LG Philips, LP064V1 LCD. The measurement data are shown in Figure 6b. The regression coefficients are thus determined as:

*a*=0.02449, *b*= –0.04984, and *c*=0.993.

The power consumption of a normally white TFT-LCD panel decreases slightly as its global transmittance increases. In other words, while maintaining the same luminance, the power consumption of the TFT-LCD decreases when dimming the backlight (which diminishes the power savings of a backlight dimming approach.).

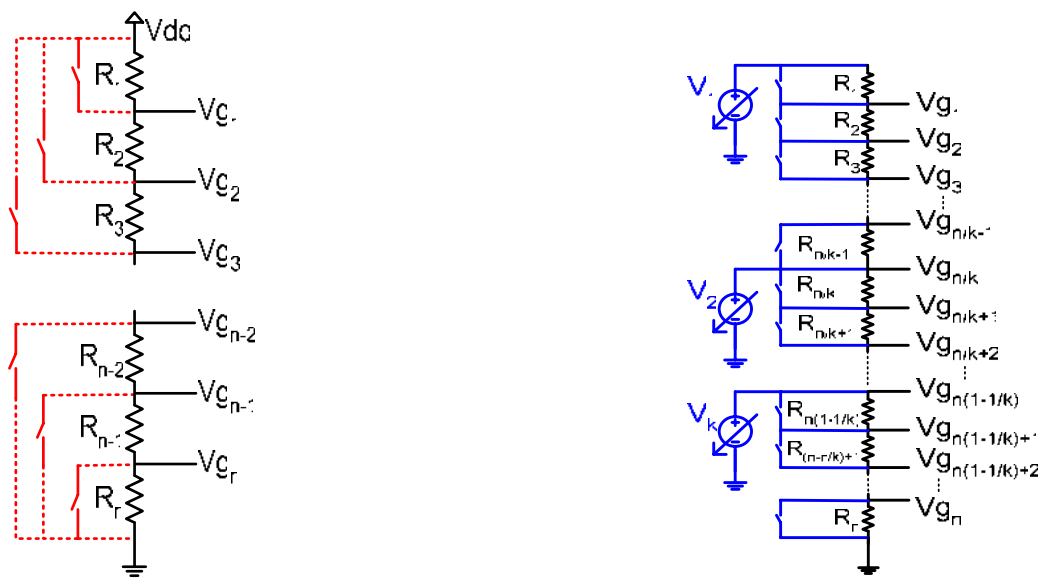

a. Conventional circuit b. Proposed circuit

Figure 5. Reference voltage divider schematics

However, the change in the TFT-LCD power consumption is quite small compared to the change in CCFL power consumption. In contrast, power consumption of normally black TFT-LCD panel increases slightly as its global transmittance increases (which increases power savings of a backlight dimming approach.) However, for either type of the TFT-LCD, the change in power consumption as a function of the transmittance is so small that it can be ignored.

**c) Distortion characteristic curve**

To characterize the distortion of different images with respect to change in their dynamic range, we set the dynamic range of a benchmark image to some target value and plot the distortion value of the transformed image as a function of this target range. We adopted the *universal image quality index* proposed in [8] as our distortion measure and used a set of benchmark images from the USC SIPI Image Database (USID) [16]. The USID is considered the de facto benchmark suite in the signal and image processing research field [9]. Figure 7 depicts the resulting distortion values for these images when the dynamic range of the transformed image is set to ten different values. Figure 8 is a subset of benchmarks reported to provide a visual reference for the distortion measure. Next, we used standard curve fitting tools provided in MATLAB version 7 release 14 to find the best "average" and "worst-case" global fits to these distortion values. The result is an empirical curve depicted in Figure 7, which maps the observed distortion function values to target dynamic range of transformed images.

### 5.2 Backlight scaling results

To show the effectiveness of HEBS approach the power saving for different images from USC SIPI database is reported in table 1. These power savings are generated for three different values of distortion levels. Clearly, by increasing the maximum tolerable distortion level the power saving should increase, which is also confirmed with listed results. Please note that the average power saving of 58% is achieved only for mere distortion level of 10%. This is about 15% increase in power saving comparing to the results reported in [4] and [5].

## 6 Conclusions

In this paper, histogram equalization for backlight scaling with pre-specified image distortion level was proposed. The proposed approach was based on an accurate definition of the image distortion. Experimental results showed the effectiveness of HEBS method. In future work alternative distortion measures and histograms equalization methods will be evaluated.

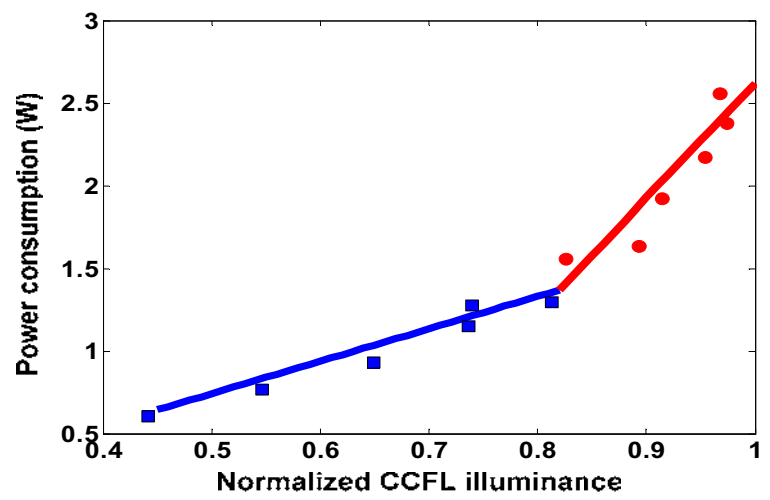

Figure 6.a. CCFL illuminance (i.e., backlight factor b) versus driver's power consumption for a CCFL source

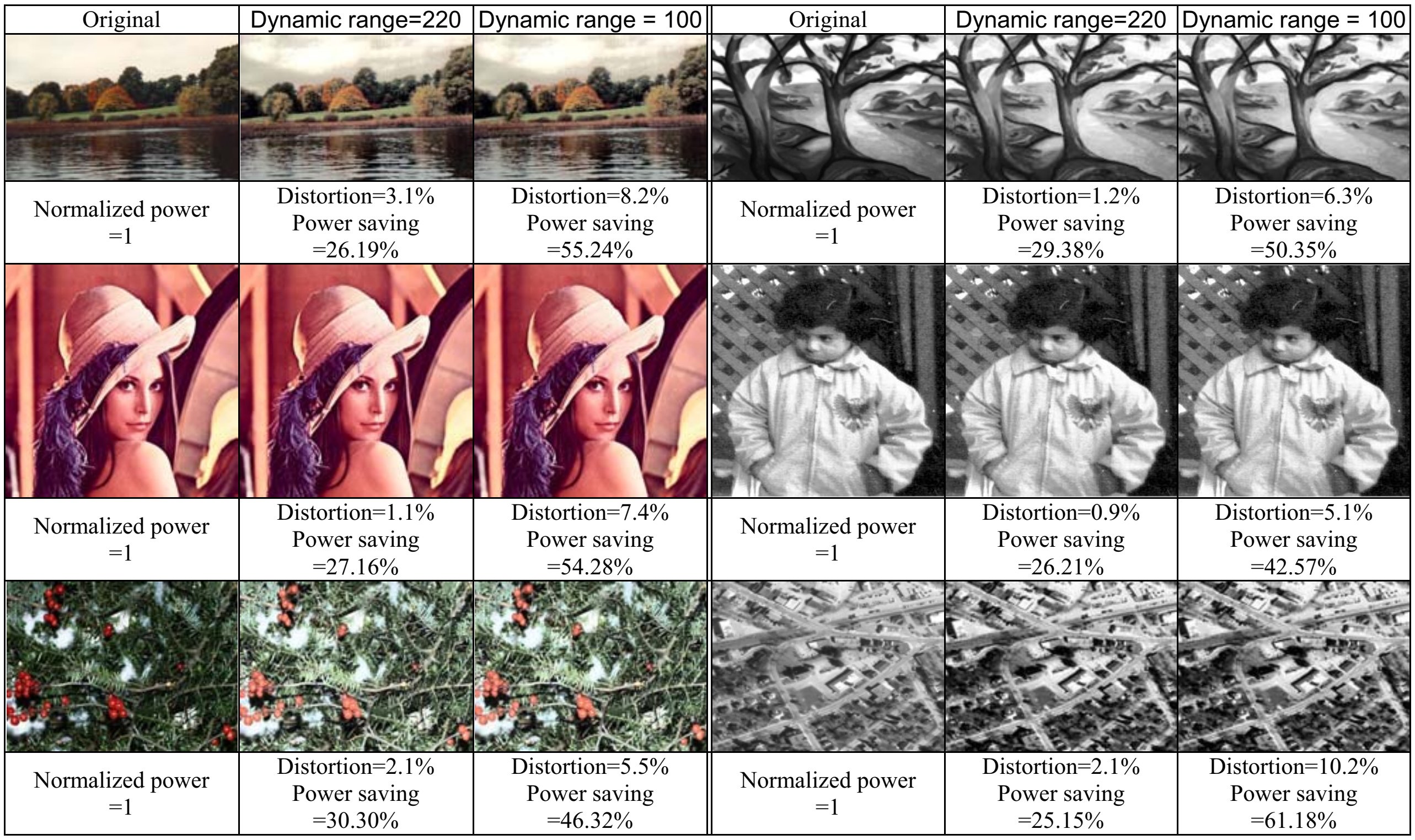

Figure 8.Sample images and their corresponding transformed versions

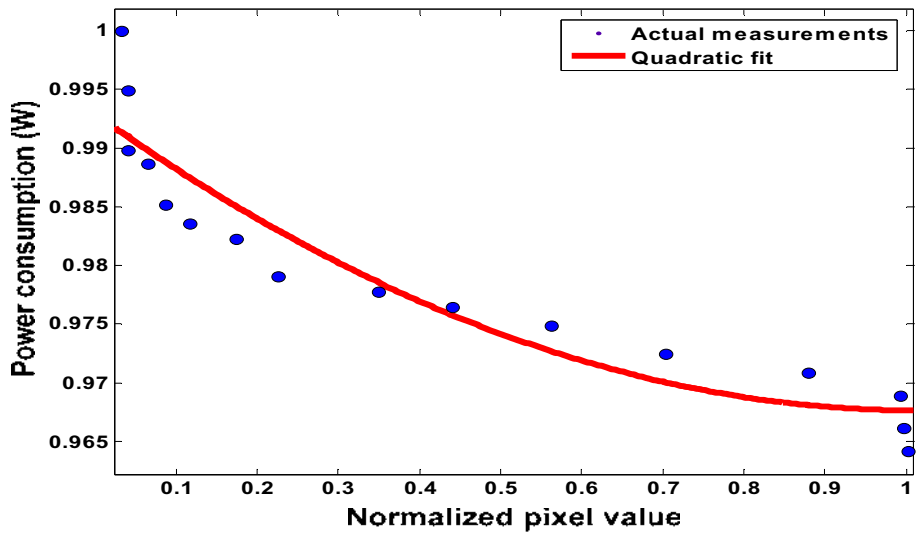

Figure 6.b. Pixel transmittance, versus power consumption of LCD panel in the normally white TFT-LCD panel

| | Power saving (%) | | |
|---|---|---|---|
| **Name** | **Distortion = 5%** | **Distortion = 10%** | **Distortion = 20%** |
| Lena | 47.53 | 58.18 | 69.52 |
| Autumn | 45.56 | 59.20 | 71.53 |
| football | 46.62 | 55.25 | 65.57 |
| Peppers | 44.60 | 54.24 | 66.55 |
| Greens | 45.63 | 55.26 | 63.58 |
| Pears | 47.51 | 57.16 | 64.49 |
| Onion | 44.56 | 58.21 | 70.53 |
| Trees | 46.69 | 54.31 | 64.62 |
| West | 48.52 | 61.18 | 67.50 |
| Pout | 42.57 | 53.22 | 59.54 |
| Sail | 42.53 | 49.18 | 56.51 |
| Splash | 46.55 | 57.20 | 63.53 |
| Girl | 46.55 | 55.20 | 62.52 |
| Baboon | 49.52 | 56.1 | 62.51 |
| TreeA | 41.53 | 50.18 | 59.52 |
| HouseA | 45.49 | 58.15 | 63.48 |
| GirlB | 45.65 | 61.28 | 62.59 |
| Testpat | 47.53 | 58.22 | 63.54 |
| Elaine | 46.53 | 55.18 | 65.50 |
| **Average** | 45.88 | 56.16 | 64.38 |

Table 1. Power saving for different distortion levels

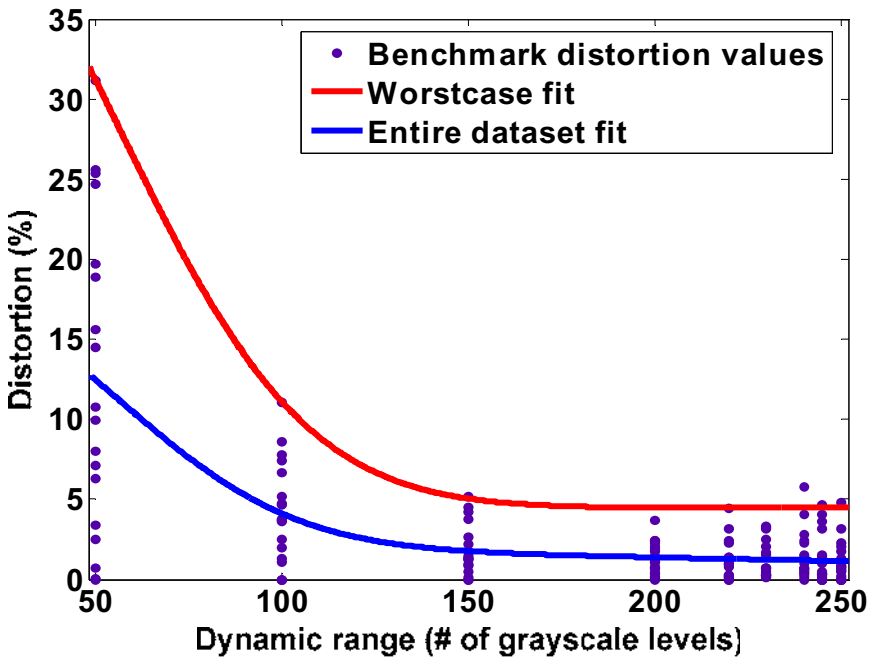

Figure 7. Distortion vs. dynamic range